\documentclass[11pt,twoside]{article}

\usepackage{xcolor}

\usepackage{fullpage}

\usepackage{epsf}
\usepackage{fancyhdr}
\usepackage{graphics}
\usepackage{graphicx}
\usepackage{psfrag}
\usepackage{comment}

\usepackage[linesnumbered,ruled]{algorithm2e}
\DontPrintSemicolon	

\usepackage{color}
\usepackage{amsthm}
\usepackage{amsfonts}
\usepackage{amsmath}
\usepackage{bm}
\usepackage{amssymb,bbm}
\usepackage{natbib}
\usepackage{algorithmic}

\usepackage{url}
\usepackage[colorlinks=True,linkcolor=magenta,citecolor=blue,urlcolor=blue,pagebackref=true,backref=true]
{hyperref}
\renewcommand*{\backref}[1]{\ifx#1\relax \else Page #1 \fi}
\renewcommand*{\backrefalt}[4]{%
    \ifcase #1 \footnotesize{(Not cited.)}%
    \or        \footnotesize{(Cited on page~#2.)}%
    \else      \footnotesize{(Cited on pages~#2.)}%
    \fi}

\usepackage{nicefrac}

\def\half{\hbox{$1\over2$}}
\def\qart{\hbox{$1\over4$}}

\usepackage{chngpage}

 \usepackage{tabularx}%

\usepackage{enumitem}
\usepackage{booktabs}
\usepackage{pbox}

\usepackage{caption,subcaption}

\usepackage{mathtools}

\usepackage{fullpage}

\newcommand{\brackets}[1]{\left[ #1 \right]}
\newcommand{\parenth}[1]{\left( #1 \right)}

\newcommand{\abss}[1]{\left| #1 \right |}

\newcommand{\Exs}{\ensuremath{{\mathbb{E}}}}

\newtheoremstyle{named}{}{}{\itshape}{}{\bfseries}{.}{.5em}{\thmnote{#3's }#1}
\theoremstyle{named}

\theoremstyle{plain}

\newtheorem{theorem}{Theorem}

\newtheorem{corollary}{Corollary}
\newtheorem{definition}{Definition}

\newtheorem{example}{Example}

\newlength{\widebarargwidth}
\newlength{\widebarargheight}
\newlength{\widebarargdepth}

\makeatletter
\long\def\@makecaption#1#2{
        \vskip 0.8ex
        \setbox\@tempboxa\hbox{\small {\bf #1:} #2}
        \parindent 1.5em  
        \dimen0=\hsize
        \advance\dimen0 by -3em
        \ifdim \wd\@tempboxa >\dimen0
                \hbox to \hsize{
                        \parindent 0em
                        \hfil
                        \parbox{\dimen0}{\def\baselinestretch{0.96}\small
                                {\bf #1.} #2
                                }
                        \hfil}
        \else \hbox to \hsize{\hfil \box\@tempboxa \hfil}
        \fi
        }
\makeatother

\long\def\comment#1{}
\definecolor{battleshipgrey}{rgb}{0.52, 0.52, 0.51}
\definecolor{darkgray}{rgb}{0.66, 0.66, 0.66}
\definecolor{darkgreen}{rgb}{0.0, 0.2, 0.13}
\definecolor{darkspringgreen}{rgb}{0.09, 0.45, 0.27}
\definecolor{dukeblue}{rgb}{0.0, 0.0, 0.61}
\definecolor{olivedrab7}{rgb}{0.24, 0.2, 0.12}
\definecolor{darkblue}{rgb}{0.0, 0.0, 0.55}
\definecolor{darkscarlet}{rgb}{0.34, 0.01, 0.1}
\definecolor{candyapplered}{rgb}{1.0, 0.03, 0.0}
\definecolor{ao(english)}{rgb}{0.0, 0.5, 0.0}
\definecolor{applegreen}{rgb}{0.55, 0.71, 0.0}

\begin{document}
\begin{center}

{\bf{\LARGE{On Integral Theorems and their Statistical Properties}}}
  
\vspace*{.2in}
{\large{
\begin{tabular}{cc}
Nhat Ho$^{\diamond}$ & Stephen G. Walker$^{\diamond, \flat}$ \\
\end{tabular}
}}

\vspace*{.2in}

\begin{tabular}{c}
Department of Statistics and Data Sciences, University of Texas at Austin$^\diamond$, \\
Department of Mathematics, University of Texas at Austin$^\flat$ \\
\end{tabular}

\date{}


\vspace*{.2in}

\begin{abstract}
We introduce a class of integral theorems based on cyclic functions and Riemann sums approximating integrals. The Fourier integral theorem, derived as a combination of a transform and inverse transform, arises as a special case. The integral theorems provide natural estimators of density functions via Monte Carlo methods. Assessments of the quality of the density estimators can be used to obtain optimal cyclic functions, alternatives to the sin function, which minimize square integrals. Our proof techniques rely on a variational approach in ordinary differential equations and the Cauchy residue theorem in complex analysis.
\end{abstract}
\end{center}

\vspace{0.2in}


\section{Introduction}
The Fourier integral theorem, see for example~\cite{Wiener33} and~\cite{Bochner_1959}, is a remarkable result. For all real integrable and continuous function $m:\mathbb{R}^{d} \to \mathbb{R}$, it yields
\begin{equation}\label{fit}
m(y)=\frac{1}{(2\pi)^{d}}\int_{\mathbb{R}^{d}}\int_{\mathbb{R}^{d}}\,\cos(s^{\top}(y-x))\,m(x)\,d x\,ds.
\end{equation}
The derivation of this result is from a combination of the Fourier and inverse Fourier transforms. As far as we are aware, it is only the cos function which yields such an integral theorem, a point supported by the paper \cite{Fowler1921}. 

If we write equation~(\ref{fit}) as
$$m(y)=\int_{\mathbb{R}^d} \psi(y-x)\,m(x)\,d x,$$
we see that we have the opportunity for obtaining a natural Monte Carlo estimator of $m(y)$ using a sample from $m(\cdot)$, provided it is integrable on $\mathbb{R}^d$.
The relevance of the 
Fourier integral theorem towards Monte Carlo estimators was observed by~\cite{Ho21}, and used in several statistics and machine learning applications, such as multivariate density estimation, nonparametric mode clustering and modal regression, quantile regression, and generative model. The methodological benefits of the Fourier integral theorem come from rewriting equation~(\ref{fit}) as
\begin{align}
m(y)=\lim_{R\to\infty} \frac{1}{\pi^{d}}\int_{\mathbb{R}^{d}} \prod_{j = 1}^{d} \frac{\sin(R(y_{j}-x_{j}))}{y_{j}-x_{j}}\,m(x)\, d x, \label{eqiv_fit}
\end{align}
where $y = (y_{1}, \ldots, y_{d})$ and $x = (x_{1}, \ldots, x_{d})$. Equation~\eqref{eqiv_fit} contains an important insight: even though we have certain dependent structures in $m(x)$, by taking the products of (independent) sin functions, the Monte Carlo estimators  are sable to capture the dependent structure. This eliminates the cumbersome and delicate procedure of choosing a covariance matrix to guarantee good practical performance of estimators based on, e.g., multivariate Gaussian kernels~\citep{Wand92, Stanis93, Chacon18}.
Further, it clearly extends the theoretical work done in the one dimensional case by \cite{Parzen62} and \cite{Davis75}.

\begin{figure}[!t]
\begin{center}
\includegraphics[width=14cm,height=7cm]{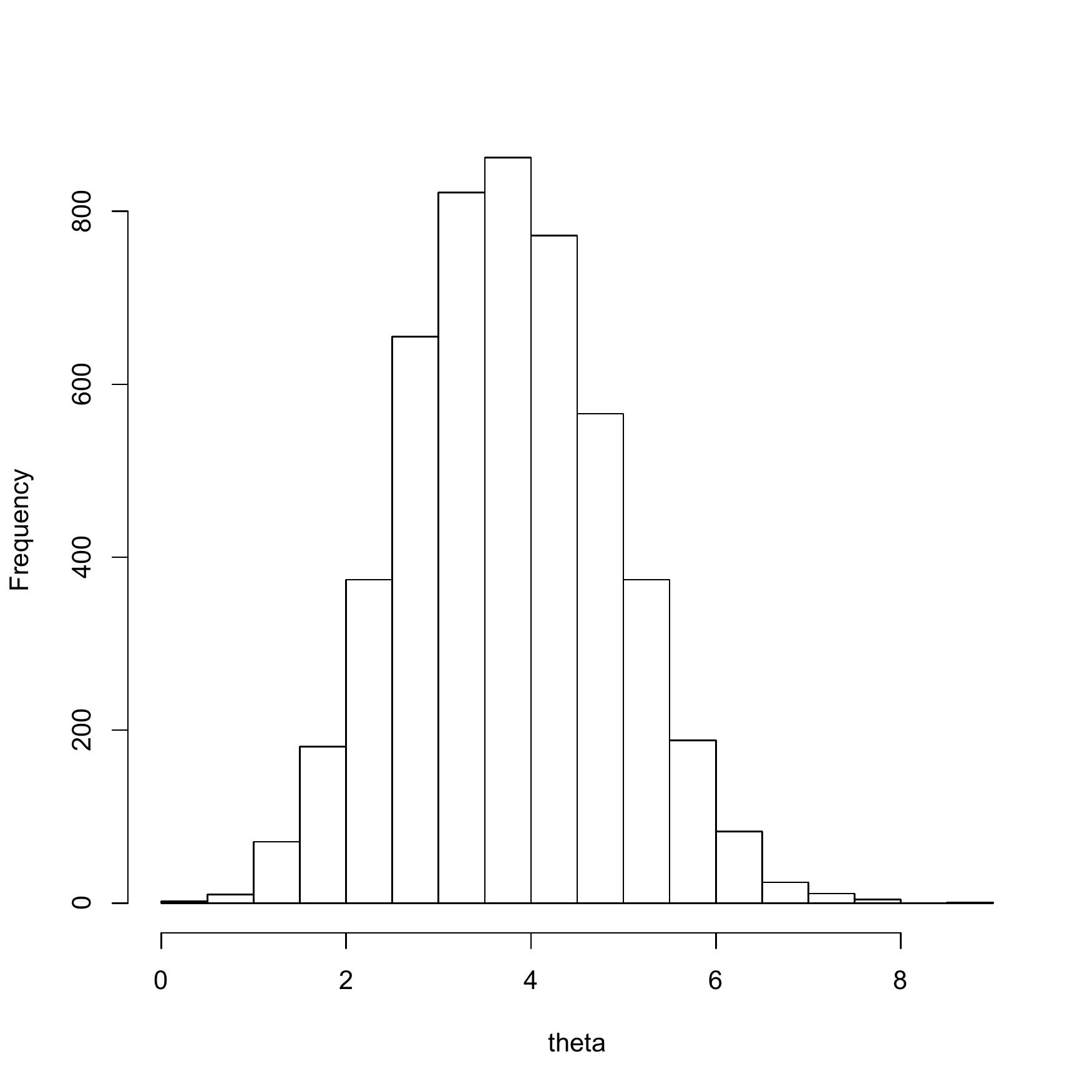}
\caption{Histogram of 1000 estimators of $\theta$; the true value is 4.}
\label{figa}
\end{center}
\end{figure}
To motivate the theory developed in the paper, here we present a statistical application for the integral theorem. We consider the density function
$$p(y\mid \theta)=\int N(y\mid 0,v^2)\,d G(v\mid 2,\theta)$$
with $\theta$ unknown and to be estimated. Here, $G$ represents the Gamma distribution. The integral to obtain a likelihood function for $\theta$ is not tractable; it is of the form
$\int v^{-1}\,e^{-\half y^2/v^2}\,v\,e^{-v\theta}\,d v.$
However, it is tractable for $y=0$. This motivates the use of the integral theorem to estimate the density at $p(0\mid\theta)$ and hence to estimate $\theta$.

Suppose $y_{1:n}$ are a sample from the model with true value $\theta_0$. Then based on the Fourier integral theorem the estimate of $p(0\mid\theta_0)$ is given by:
$$\widehat{p}(0\mid\theta_0)=n^{-1}\sum_{i=1}^n \frac{\sin(R\,y_i)}{\pi\,y_i}$$
and the correct value of $p(0\mid\theta)=\theta/\sqrt{2\pi}$. Therefore the estimator of $\theta$ is
$$\widehat{\theta}=\widehat{p}(0\mid\theta_0)\,\sqrt{2\pi}.$$
By way of illustration we took $n=100$ with $\theta_0=4$.
The choice of $R$ was 50 and we repeated the experiment 1000 times. A histogram of the 1000 estimators is presented in Fig.~\ref{figa}.
The mean value is 3.75.

Clearly this example is a simple case, and the $\theta$ could be estimated using numerical routines such as an Expectation-Maximization (EM) algorithm~\citep{Rubin-1977}.
Nevertheless, even an EM algorithm is difficult to implement in this case. Further examples of the integral theorem appear in~\cite{roti22} where it is used to compute Bayesian marginal likelihoods.

\begin{figure}[!t]
\begin{center}
\includegraphics[width=14cm,height=7cm]{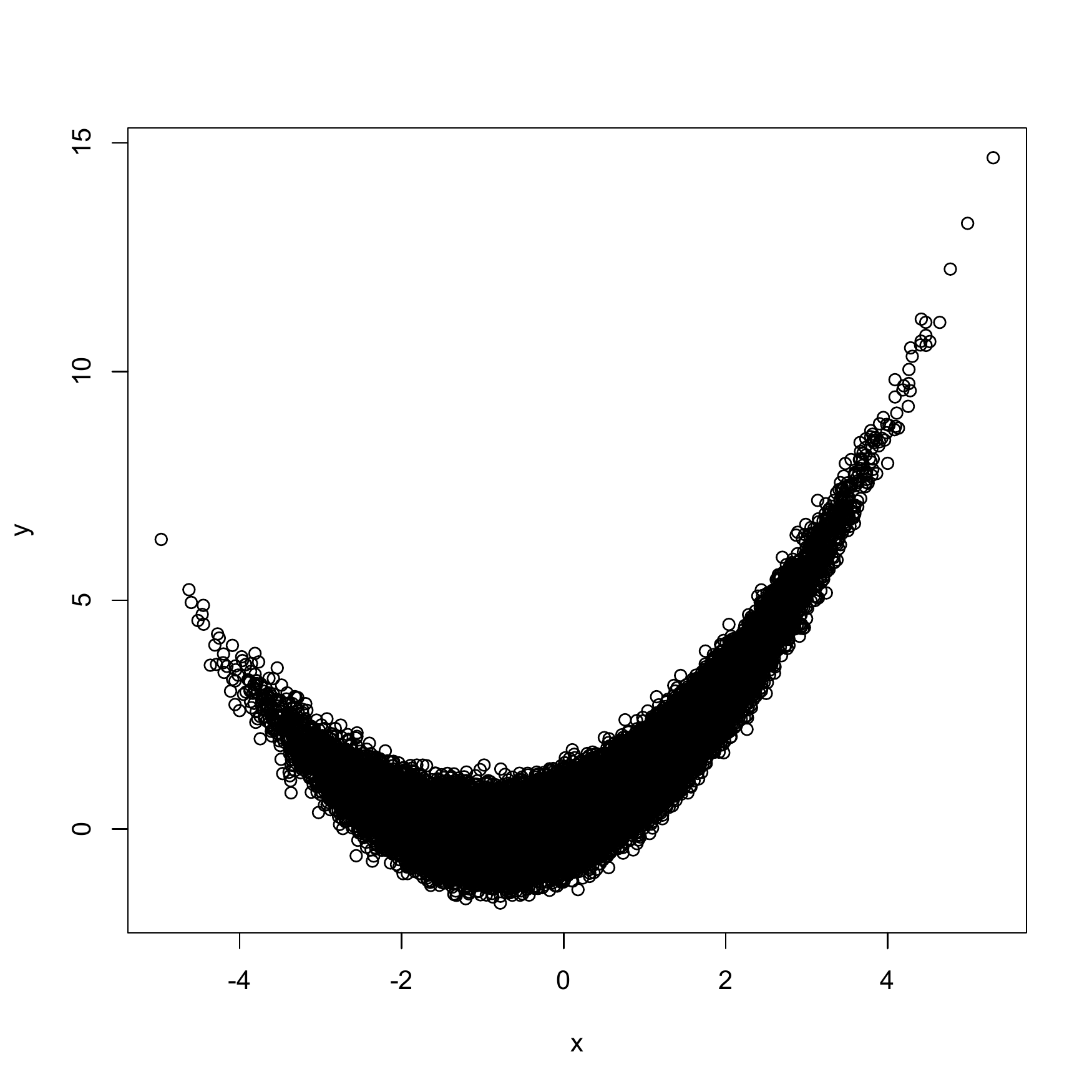}
\caption{Sample from bivariate banana shaped density.}
\label{figt1}
\end{center}
\end{figure}

\begin{figure}[!t]
\begin{center}
\includegraphics[width=14cm,height=7cm]{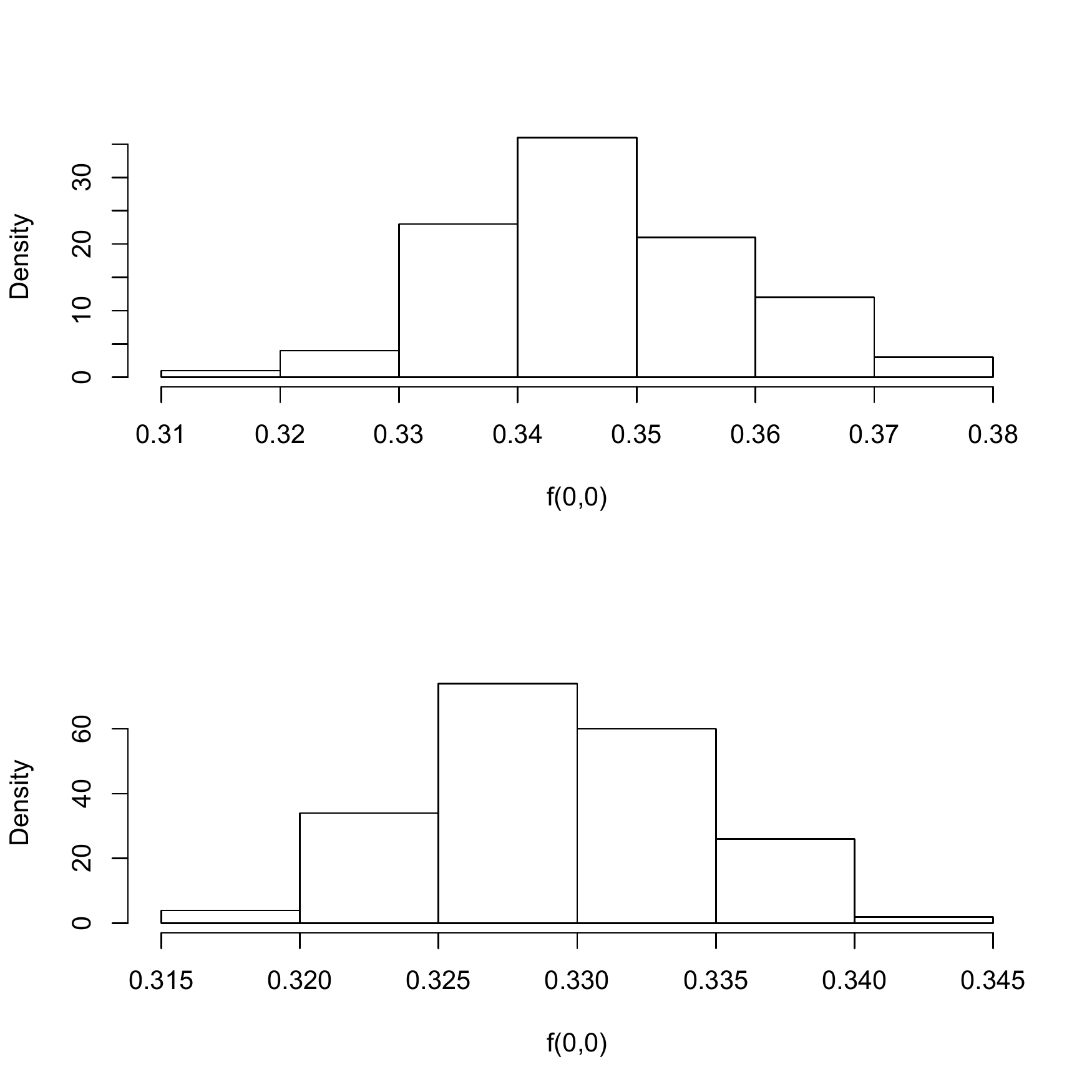}
\caption{Histogram of estimators of $f(0,0)$ using Fourier integral theorem (upper) and an independent Gaussian kernel density (lower).}
\label{figt2}
\end{center}
\end{figure}

Another demonstration of the integral theorem highlights the lack of need for a covariance matrix when estimating with a bivariate sample. We take $f(x,y)$ to be a banana shaped density; see Fig.~\ref{figt1}. We use these samples to estimate $f(0,0)$, the true value of which is 0.347.
In Fig.~\ref{figt2} we show the histogram of 100 experiments using the Fourier integral theorem (upper figure) and an independent Gaussian kernel density estimator using the recommended bandwidth for the appropriate sample size. In particular, for $n=100,000$, we take $R=20$ and the bandwidth for each Gaussian kernel to be Silverman's rule of thumb, e.g. $h_x=n^{-1/6}\,\sqrt{S_x}$, where $S_x$ is the sample variance of the $x$ values. The histogram from the Fourier integral theorem accurately picks out the true value for $f(0,0)$, in fact the mean is precisely $0.347$, whereas the kernel density estimator is clearly biased. This shows that whereas a covariance matrix is not required for the Fourier integral theorem, it is recommended for the Gaussian kernel density~\citep{WJones93, Wand92, Stanis93, Chacon18}. The reason is that the Fourier estimator is using an unbiased Monte Carlo sampler whereas the Gaussian kernel density estimator is not. 


These illustrations motivate the need for a study of not just the Fourier integral theorem, but integral theorems in general.
The aim in this paper is to introduce a general class of integral theorems to be used as Monte Carlo estimators with applications in statistics and machine learning. The reasoning is as follows; for different function $m$ there will possibly be alternative integral theorems which provide better Monte Carlo estimators. Indeed, we do find new integral theorems which provide superior Monte Carlo estimators.

We do not come at this new class of integral theorems using transforms and inverse transforms, but rather use the novel combination of two well known objects; a cyclic function which integrates to 0 over each cyclic interval, and a Riemann sum approximation to an integral. 

In particular, we define almost everywhere differentiable cyclic functions $\phi:\mathbb{R}\to\mathbb{R}$ such that
\begin{equation}\label{eq:normalized_constant}
\int_0^{2\pi} \phi(x)\,d x=0\quad\mbox{and}\quad \int_{\mathbb{R}}\frac{\phi(x) }{x}\,d x=1.
\end{equation}
The Fourier integral theorem corresponds to
\begin{align}
\phi(x)=\frac{1}{\pi}\sin x. \label{eq:Fourier_kernel}
\end{align}
Then, via Riemann sums approximating integrals, we demonstrate that
\begin{align}
m(y)=\lim_{R\to\infty} \int_{\mathbb{R}^{d}} \prod_{j = 1}^{d} \frac{\phi(R(y_{j} - x_{j}))}{y_{j} - x_{j}}\,m(x)\, dx. \label{eq:general_class_integral}
\end{align}
Similar to the Fourier integral theorem, \eqref{eqiv_fit}, the general integral theorem in equation~\eqref{eq:general_class_integral} are also able to automatically preserve the dependence structures in the function $m$. 
With our finding of large classes of integral functions, the question posed is which kernel $\phi$ has optimal properties in terms of estimation. 

In this work, we specifically answer this question in the context of kernel density estimation problem~\citep{Rosen1956, Parzen62, Epanechnikov_1969, Yakowitz_1985, Devroye_Nonparametric, Scott_Variable, WJones93, Wass06, Botev_Diffusion, Nickl_Confidence, Jiang_Uniform}. Indeed, the kernel density estimator based on equation~\eqref{eq:general_class_integral} would be given by:
\begin{align}
\widehat{m}_{R}(y)=\frac{1}{n}\sum_{i=1}^n \prod_{j = 1}^{d} \frac{\phi(R(y_{j}-X_{ij}))}{y_{j}-X_{ij}}, \label{eq:multivariate_smoothing_estimator}
\end{align}
where $X_{1}, \ldots, X_{n} \in \mathbb{R}^{d}$ represent the sample from the density function $m$ and a finite $R$ is required for the smoothing. We study upper bounds for bias of the estimator $\widehat{m}_{R}$ based on $R$ and the sample size $n$ in Theorem~\ref{theorem:approximation_Fourier} and Corollary~\ref{corollary:approx_error_general_integral}.

In order to find the optimal kernel $\phi$, we use  asymptotic mean integrated square error, which is key property determining the quality of an estimator; see for example ~\cite{Wand92, Wand94}. To ease the findings, we specifically consider the univariate settings, i.e., $d = 1$ and use the following two terms stemming from that error to determine the optimal kernel: the term
\begin{equation}\label{opt_optimistic}
\int_{\mathbb{R}} \left(\frac{\phi(x)}{x}\right)^2 \,d x,
\end{equation}
which provides an upper bound on the variance of density estimator $\widehat{m}_{R}$; and the term
\begin{equation}\label{opt_constrained}
\int_{\mathbb{R}} \left(\frac{\phi(x)}{x}\right)^2 m(x) \,d x,
\end{equation}
which yields more precise asymptotic behaviors of the variance of the density estimator $\widehat{m}_{R}$ than that from equation~\eqref{opt_optimistic}. To support the idea of only considering the variances, we show that the bias does not depend on $\phi$.

We demonstrate that by minimizing the first term~\eqref{opt_optimistic} subject to the constraints~\eqref{eq:normalized_constant}, the optimal kernel $\phi$ is the sin function~\eqref{eq:Fourier_kernel} in the Fourier integral theorem. This is achieved via a variational approach in ordinary differential equations. On the other hand, by using the Cauchy residue theorem in complex analysis, we prove that the optimal kernel is not sin kernel for minimizing the second term~\eqref{opt_constrained} subject to the constraints~\eqref{eq:normalized_constant}. It also demonstrates the usefulness in finding other integral theorems to the Fourier integral theorem.

The organization of the paper is as follows. In Section~\ref{Sec:property_Fourier}, we first revisit the Fourier integral theorem and establish the bias of its density estimator via Riemann sums approximating integrals. Then, using the insight from that theorem, we introduce a general class of integral theorems that possess similar approximation errors. After deriving our class of integral theorems, in Section~\ref{sec:optimality_sin} we study optimal kernels that minimize either the problem~\eqref{opt_optimistic} or the problem~\eqref{opt_constrained} 
subject to the constraints (\ref{eq:normalized_constant}). Finally, we conclude the paper with a few discussions in Section~\ref{sec:discussion} while deferring the proofs of the remaining results in the paper to the Appendix.
\section{Integral Theorems}
\label{Sec:property_Fourier}
We first study the bias of kernel density estimator~\eqref{eq:multivariate_smoothing_estimator} or equivalently approximation property of the Fourier integrals theorem via the Riemann sums approximating integral theorem in Section~\ref{sec:Fourier_integral}. Then, using the insight from that result, we introduce a general class of integral theorem that possesses similar approximation behavior to the Fourier integral theorem, in Section~\ref{sec:general_integral}.

\subsection{The Fourier integral theorem revisited}
\label{sec:Fourier_integral}
Before going into the details of the general integral theorem, we reconsider the approximation property of the Fourier integral theorem. In~\cite{Ho21} the authors utilize the tail behavior of the Fourier transform of the function $m(\cdot)$ to characterize an approximation error of the Fourier integral theorem when truncating one of the integrals. However, the technique in the proof is inherently based on properties of the sin kernel 
and is non-trivial to extend to other choices of cyclic function; examples of such functions are provided in Section~\ref{sec:general_integral}. 

In this paper, we provide insight into the approximation error of the Fourier integral theorem via the Riemann sum approximating to an integral. This insight can be generalized into any cyclic function which integrates to 0 over the cyclic interval, thereby enriching the family of integral theorems beyond  Fourier's. To simplify the presentation, we define
\begin{align}
    m_{R}(y) & : = 
    \frac{1}{\pi^d}\int_{\mathbb{R}^{d}} \prod_{j = 1}^{d} \frac{\sin(R(y_j-x_j))}{(y_{j} - x_{j})} m(x) \, dx. \label{eq:Fourier_approx}
\end{align}
By simple calculation, $m_{R}(y) = \mathbb{E}(\widehat{m}_{R}(y))$ where the outer expectation is taken with respect to i.i.d. samples $X_{1}, \ldots, X_{n}$ from $m$ and $\widehat{m}_{R}$ is the density estimator~\eqref{eq:multivariate_smoothing_estimator} when $\phi$ is the sin kernel. Therefore, to study the bias of the kernel density estimator $\widehat{m}_{R}$ in equation~\eqref{eq:multivariate_smoothing_estimator}, it is sufficient to consider the approximation error of the Fourier integral theorem, namely, we aim to upper bound $\abss{m_{R}(y) - m(y)}$ for all $y$. To obtain the bound, we start with the following definition of the class of univariate functions that we use throughout our study.

\vspace{0.1in}
\noindent
\begin{definition}
\label{def:class_function}
The univariate function $f(\cdot)$ is said to belong to the class $\mathcal{T}^{K}(\mathbb{R})$ if for any $y \in \mathbb{R}$, the function $g(x) = (f(x) - f(y))/(x - y)$ satisfies the following conditions:
\begin{enumerate}
    \item The function $g$ is differentiable, uniformly continuous up to the $K$-th order, and the limits $\lim_{|x| \to +\infty} | g^{(k)}(x)| = 0$ for any $0 \leq k \leq K$ where $g^{(k)}(.)$ denotes the $k$-th order derivative of $g$;
    \item The integrals $\int_{\mathbb{R}} |g^{(k)}(x)| dx$ are finite  for all $0 \leq k \leq K$.
\end{enumerate}
\end{definition}

\vspace{0.1in}
\noindent
Note that, for the function $g$ in Definition~\ref{def:class_function}, for any $y \in \mathbb{R}$ when $x = y$, we choose $g(y) = f^{(1)}(y)$. Based on Definition~\ref{def:class_function}, we now state the following result.
\begin{theorem}
\label{theorem:approximation_Fourier}
Assume that the univariate functions $m_{j} \in \mathcal{T}^{K_{j}}(\mathbb{R})$ for any $1 \leq j \leq d$ where $K_{1}, \ldots, K_{d}$ are given positive integer numbers. Then, if we have $m(x) = \prod_{j = 1}^{d} m_{j}(x_{j})$ or $m(x) = \sum_{j = 1}^{d} m_{j}(x_{j})$ for any $x = (x_{1}, \ldots, x_{d})$, there exist universal constants $C$ and $\bar{C}$ depending on $d$ such that as long as $R \geq C$ we obtain
\begin{align*}
    \abss{m_{R}(y) - m(y)} \leq \bar{C}/R^{K},
\end{align*}
where $K = \min_{1 \leq j \leq d} \{K_{j}\}$.
\end{theorem}
\noindent
The proof is presented in the Appendix. To appreciate the proof we demonstrate the key idea in the one dimensional case. Here
$$m_R(y)-m(y)=\frac{1}{\pi}\int_{-\infty}^{+\infty} \frac{\sin(R(y-x))}{y-x}\,(m(x)-m(y))\,d x$$
which we write as 
$$m_R(y)-m(y)=\frac{1}{\pi}\int_{-\infty}^{+\infty} \sin(R(x-y))\,g(x)\,d x,$$
where $g(x)=(m(x)-m(y))/(x-y)$. Without loss of generality, we set $y=0$ to get
$$m_R(y)-m(y)=\frac{1}{\pi}\int_{-\infty}^{+\infty} \sin(z)\,\epsilon g(z\epsilon)\,d z,$$
where $\epsilon=1/R$. Now due to the cyclic behaviour of the sin function we can write this as
$$m_R(y)-m(y)=\frac{1}{\pi}\int_{0}^{2\pi} \sin (t)\,\sum_{k=-\infty}^{+\infty} \epsilon g(\epsilon(t+2\pi k))\,\,d t.$$
The term
$\sum_{k=-\infty}^{+\infty} \epsilon g(\epsilon(t+2\pi k))$
is a Riemann sum approximation to an integral which converges to a constant, for all $t$, as $\epsilon \to 0$.
The overall convergence to 0 is then a consequence of $\int_0^{2\pi} \sin\,t \,d t=0$. Hence, it is how the Riemann sum converges to a constant which determines the speed at which $m_R(y)-m(y)\to 0$.

\subsection{General integral theorem}
\label{sec:general_integral}
It is interesting to note that the sin function in the Fourier integral theorem could be replaced by any cyclic function which integrates to 0 over the cyclic interval. In particular, we consider the following general form of integral theorem:
\begin{align}
	m_{R, \phi}(y) = \int_{\mathbb{R}^{d}} \prod_{j = 1}^{d} \frac{\phi(R(y_{j} - x_{j}))}{(y_{j} - x_{j})} m(x) dx, \label{eq:general_integral_theorem}
\end{align}
where the univariate function $\phi$ is a cyclic function on $(0, 2 \pi)$. 
Using the proof technique of Theorem~\ref{theorem:approximation_Fourier} and the assumptions with function $m$ in that theorem, we also obtain the following approximation error of the general integral theorem:
\begin{corollary}
\label{corollary:approx_error_general_integral} Assume that the kernel $\phi$ satisfies the constraints~\eqref{eq:normalized_constant} and the function $m$ satisfies the assumptions in Theorem~\ref{theorem:approximation_Fourier}. Then, there exist universal constants $C$ and $\bar{C}$ depending on $d$ and the function $\phi$ such that when $R \geq C$ we have
\begin{align*}
    \abss{m_{R,\phi}(y) - m(y)} \leq C/ R^{K}
\end{align*}
where $K$ is defined as in Theorem~\ref{theorem:approximation_Fourier}.
\end{corollary} 
\noindent
Therefore, we have a general class of integral theorems that possesses similar approximation errors as that of the Fourier integral theorem. Furthermore, the general integral theorems are also able to automatically maintain the dependence structures in function $m$.

We now discuss some examples of function $\phi$ that have connection to Haar wavelet and splines.
\begin{example} (Haar wavelet integral theorem) We consider the piece-wise linear function
\begin{align*}
	\phi_{\text{Haar}}(x) = \begin{cases} \dfrac{(x - 2k \pi)2}{C_{1} \pi}, \ \text{if} \ x \in ((2k - \frac{1}{2}) \pi, (2k + \frac{1}{2})\pi)) \\ \dfrac{((2k+1)\pi - x)2}{C_{1} \pi}, \ \text{if} \ x \in ((2k + \frac{1}{2}) \pi, (2k + \frac{3}{2})\pi) \end{cases}
\end{align*}
where $$C_{1} = \sum_{k = -\infty}^{\infty} \parenth{2(2k+1) \log \parenth{\frac{4k+3}{4k+1}} - 4k \log \parenth{\frac{4k+1}{4k-1}}},$$ 
and 
$$\phi_{\text{Haar}}'(x) = 
\left\{\begin{array}{ll}
\frac{2}{C_{1} \pi} & x \in ((2k - \frac{1}{2}) \pi, (2k + \frac{1}{2})\pi)) \\
 - \frac{2}{C_{1} \pi} & x \in ((2k + \frac{1}{2}) \pi, (2k + \frac{3}{2})\pi)).
 \end{array}\right.
$$
It demonstrates that the derivative of the function $\phi$ is the Haar wavelet function, which can be named the ``Haar wavelet integral theorem" for this choice of $\phi_{\text{Haar}}$. 
\end{example}
\begin{example} (Spline integral theorem) Here we take into account the piece-wise quadratic function
\begin{align*}
	\phi_{\text{spline}}(x) = \begin{cases} \dfrac{4(x - 2k \pi)(x - (2k - 1)\pi)}{C_{2} \pi^2}, \ \text{if} \ x \in ((2k - 1) \pi, 2k \pi) \\
	\dfrac{4(x - 2k \pi)((2k+1)\pi - x)}{C_{2} \pi^2}, \ \text{if} \ x \in (2k  \pi, (2k + 1)\pi) \end{cases}
\end{align*}
where 
$$C_{2} = \sum_{k = -\infty}^{\infty} \parenth{8k(2k-1)\log \parenth{\frac{2k}{2k-1}} - 8k(2k+1)\log \parenth{\frac{2k + 1}{2k}}}.$$ 
A direct calculation shows that 
$$\phi_{\text{spline}}'(x) = \left\{\begin{array}{ll}
\frac{8x - 4(4k - 1) \pi}{C_{2} \pi^2} &  x \in ((2k - 1) \pi, 2k \pi) \\ 
\frac{-8x + 4(4k + 1) \pi}{C_{2} \pi^2} &  x \in (2k \pi, (2k+1) \pi).
\end{array}\right.$$
Therefore, the first derivative of $\phi_{\text{spline}}$ is a piece-wise linear function. The particular form of $\phi_{\text{spline}}'$ justifies the spline integral theorem for this choice of kernel function $\phi_{\text{spline}}$.
\end{example}
Finally, we would like to highlight that the Haar wavelet and spline integral theorems are just two instances of the integral theorems. In general, the class of cyclic function $\phi$ satisfying an integral theorem is vast.
\section{Optimal Functions}
\label{sec:optimality_sin}
In this section, we discuss optimal functions $\phi$ from the integral theorem with respect to the kernel density estimation problem. To ease the findings, we specifically consider the univariate settings, namely, $d = 1$. Subject to the constraints in equation~\eqref{eq:normalized_constant}, as mentioned in the introduction, we consider minimizing either the problem~\eqref{opt_optimistic} or the problem~\eqref{opt_constrained}. We show that the sin function minimizes
$$\int_\mathbb{R} \left(\frac{\phi(x)}{x}\right)^2\,d x$$
subject to constraints in equation~\eqref{eq:normalized_constant}, whereas this is not the case when we introduce a density function, i.e., the aim now being to minimize
$$\int_\mathbb{R} \left(\frac{\phi(x)}{x}\right)^2\,m(x)\,d x$$
for some density function $m$, which yields more precise asymptotic behaviors of the variance of density estimator $\widehat{m}_{R}$ in equation~\eqref{eq:multivariate_smoothing_estimator}.

Before proceeding we briefly explain why the bias is not relevant, as it only depends on $m$. To see this we can write the bias as
$$R^{-1}\,\int \phi(s)\,\left\{\frac{m(y+s/R)-m(y)}{s/R}\right\}\,d s.$$ 
How this goes to zero as $R\to\infty$ depends solely on the $(m(y+s/R)-m(y))/s$ term and how it converges to $m'(y)$; recall the $\phi$ is cyclic and integrates to zero over every interval of the type $(2\pi k,2\pi(k+1))$. 

\subsection{The sin function }
As we have mentioned, a direct application of the general integral theorem is for a Monte Carlo estimator of density functions;~\cite{Ho21}. 
The bias of the estimator $\widehat{m}_{R}$, namely, $\Exs \brackets{\widehat{m}_{R}(y)} - m_{R}(y)$, has been established in Corollary~\ref{corollary:approx_error_general_integral}. A natural question to ask is the form of optimal kernel $\phi$ that leads to a good quality of density estimator $\widehat{m}_{R}$. To answer that question, we use asymptotic mean integrated square error, see~\cite{Wand92} and \cite{Wand94}, which is equivalent to find the optimal kernel $\phi$ which leads to a small variance for the estimator $\widehat{m}_{R}$. A simple calculation shows that
\begin{align}
    \text{Var}(\widehat{m}_{R}(y)) = \frac{1}{n} \text{Var} \parenth{\frac{\phi(R(y - X))}{(y - X)}}, \label{eq:variance}
\end{align}
for any $y \in \mathbb{R}$ where the outer variance is taken with respect to random variable $X$ following density function $m$. With an assumption that $\|m\|_{\infty} < \infty$, we can upper bound the variance of $\widehat{m}_{R}(y)$ as follows:
\begin{align}
    \text{Var}(\widehat{m}_{R}(y)) \leq \frac{R \|m\|_{\infty}}{n}  \int_{-\infty}^{\infty} \frac{\phi^2(x)}{x^2} dx. \label{eq:variance_upper_bound}
\end{align}
The integral in the upper bound in equation~\eqref{eq:variance_upper_bound} is convenient as it involves $m$. It indicates that the optimal kernel $\phi$ minimizing that integral will be independent of $m$, which also yields good insight into the behavior of the variance of the density estimator $\widehat{m}_{R}$ for all $y$ and $m$. Therefore, we consider minimizing the upper bound~\eqref{eq:variance_upper_bound} with respect to the constraints that $\phi$ is almost surely differentiable cyclic function in $(0, 2\pi)$ and satisfies the constraints~\eqref{eq:normalized_constant}. It is equivalent to
solving the following objective function
\begin{equation*}
	 \min_{\phi} \int_{-\infty}^{\infty} \frac{\phi^2(x)}{x^2} dx,
\end{equation*}	 
such that $\phi$ satisfies \eqref{eq:normalized_constant}. This is the objective function~\eqref{opt_optimistic}  mentioned in the introduction.

To study the optimal function $\phi$ that satisfy these constraints, we define the following functions:
\begin{align*}
	\alpha(x) : = \sum_{k = -\infty}^{\infty} \frac{1}{x + 2k\pi}, \quad \beta(x)  : = \sum_{k = -\infty}^{\infty} \frac{1}{(x + 2k\pi)^2}
\end{align*}
for any $x \in (0, 2\pi)$. Now, we would like to prove that
\begin{align}
    \alpha(x) = \frac{1}{2 \tan(x/2)}, \quad \beta(x) = -\frac{1}{2 \sin^2(x/2)}. \label{eq:key_equations}
\end{align}
In fact, from the infinite product representation of the sin function, we have
\begin{align}
    \frac{\sin(\pi t)}{\pi t} = \prod_{k= 1}^{\infty} \parenth{1 - \frac{t^2}{k^2}}, \label{eq:sinc_represent}
\end{align}
for any $t \in (0,1)$. By taking the logarithm of both sides of the equation and take the derivative with respect to $x$, we obtain that
\begin{align*}
    \frac{\pi \cos(\pi t)}{\sin(\pi t)} - \frac{1}{t} = - \sum_{k = 1}^{\infty} \frac{2 t}{k^2 - t^2}.
\end{align*}
By using the change of variable changing $t = x/(2\pi)$, we obtain the conclusion that $2 \alpha(x) = \cot(x/ 2)$. The form of $\beta(x)$ can be obtained direct by taking the derivative of $\alpha(x)$. Therefore, we obtain the conclusion of claim~\eqref{eq:key_equations}.

Now, we state our main result for the optimal kernel $\phi$ solving the objective function~\eqref{opt_optimistic}.  
\begin{theorem}
\label{theorem:optimal_function}
The optimal cyclic and almost everywhere differentiable function $\phi$ that solves the objective function~\eqref{opt_optimistic} subject to constraints~\eqref{eq:normalized_constant} is $\phi(x) = \sin(x)/ \pi$ for all $x \in (0, 2\pi)$.
\end{theorem}
\noindent
Interestingly, if we consider a truncation of the sin function, which corresponds to the optimal kernel $\phi$ for solving objective function~\eqref{opt_optimistic}, at $k = 1$ in equation~\eqref{eq:sinc_represent}, we obtain the Epanechnikov kernel,~\cite{Epanechnikov_1969} and \cite{Mueller1984}, which is given by
    $k_{\text{Epa}}(x) = 3(1 - x^2)/4$, 
for $x \in (-1, 1)$ and 0 otherwise. This kernel had been shown to have optimal efficiency among non-negative kernels that are differentiable up to the second order; ~\cite{Tsy09}. Direct calculation shows that
\begin{align*}
    \int_{\mathbb{R}} k_{\text{Epa}}^2(x) dx = \frac{3}{5} > \int_{\mathbb{R}} \frac{\sin^2(x)}{\pi^2 x^2}dx = \frac{1}{\pi}.
\end{align*}
Therefore, if we use the term~\eqref{opt_optimistic} as an indication for the quality of our variance, the Epanechikov kernel is not better than the sin kernel from the Fourier integral theorem. It also aligns with an observation from~\cite{Tsy09} that we can construct better kernels, which can take negative values, than the Epanechnikov kernel without restricting to only the non-negative kernels.
\begin{proof} Since the function $\phi$ is cyclic in $(0, 2\pi)$, we have
\begin{align*}
	\int_{-\infty}^{\infty} \frac{\phi(x))}{x} dx = \sum_{k = -\infty}^{\infty} \int_{2k \pi}^{2(k+1)\pi} \frac{\phi(x)}{x} dx =  \sum_{k = -\infty}^{\infty} \int_{0}^{2 \pi} \frac{\phi(x + 2k \pi)}{x + 2k \pi} dx = \int_{0}^{2 \pi} \phi(x) \alpha(x) dx.
\end{align*}
Similarly, we also obtain that
\begin{align*}
	\int_{-\infty}^{\infty} \frac{\phi^2(x)}{x^2} dx = \int_{0}^{2\pi} \phi^2(x) \beta(x) dx. 
\end{align*}
Given the above equations, the original 
problem can be rewritten as follows:
\begin{align}
	& \min_{\phi} \int_{0}^{2 \pi} \phi^2(x) \beta(x) dx, \label{eq:equivalent_variance} \\
	\text{such that} & \int_{0}^{2\pi} \phi(x) dx = 0, \ \int_{0}^{2\pi} \phi(x) \alpha(x) d x = 1. \nonumber
\end{align} 
The Lagrangian function corresponding to the objective function~\eqref{eq:equivalent_variance} takes the form:
\begin{align*}
	\mathcal{L}(\phi) = \int_{0}^{2 \pi} \phi^2(x) \beta(x) dx - \lambda_{1} \int_{0}^{2\pi} \phi(x) dx - \lambda_{2} \left(\int_{0}^{2\pi} \phi(x) \alpha(x) d x  - 1\right).
\end{align*}
Since $\phi$ is almost surely differentiable, the function $\mathcal{L}$ can be rewritten as follows:
\begin{align*}
	\mathcal{L}(\phi) & = \phi^2(2\pi) B(2\pi) - \phi^2(0) B(0) - \int_{0}^{2 \pi} 2 \phi(x) \phi'(x) B(x) dx - \lambda_{1} \int_{0}^{2\pi} \phi(x) dx \\
	& \hspace{20 em} - \lambda_{2} \left(\int_{0}^{2\pi} \phi(x) \alpha(x) d x  - 1\right) \\
	& = - \int_{0}^{2\pi} \phi(x) \biggr(2 \phi'(x) B(x) + \lambda_{1} + \lambda_{2} \alpha(x)\biggr)dx - C \\
	& : = \int_{0}^{2\pi} F(x) dx - C, 
\end{align*}
where $B$ is the function such that $B'(x) = \beta(x)$ and $C = \lambda_{2} + \phi^2(2\pi) B(2\pi) - \phi^2(0) B(0)$. To find $\phi$ that minimizes the function $\mathcal{L}$, we use the Euler-Lagrange equation, see for example~\cite{Young1969}, which entails that
\begin{align*}
	\frac{\partial{F}}{\partial{\phi}} (\phi) - \frac{\partial}{\partial{x}} \biggr(\frac{\partial{F}}{\partial{\phi'}} (\phi)\biggr) = 0.
\end{align*}
This equation leads to 
\begin{align*}
	\phi(x) = \frac{-\lambda_{1} - \lambda_{2} \alpha(x)}{2\beta(x)},
\end{align*}
where $\lambda_{1}$ and $\lambda_{2}$ can be determined by solving the conditions~\eqref{eq:normalized_constant}. 

Given the form of optimal $\phi$ and the forms of $\alpha(x)$ and $\beta(x)$ in equation~\eqref{eq:key_equation}, the first condition in equation~\eqref{eq:normalized_constant} leads to
\begin{align*}
    \int_{0}^{2\pi} \parenth{2\lambda_{1} \sin^2(x/2) + \lambda_{2} \sin(x/2) \cos(x/2)} dx = 0. 
\end{align*}
This yields $\lambda_{1} = 0$. 
The second condition in equation~\eqref{eq:normalized_constant} indicates that
\begin{align*}
    \int_{0}^{2\pi} \lambda_{2} \cos^2(x/2) dx = 4,
\end{align*}
which leads to $\lambda_{2} = \pi/4$. Therefore, we have the optimal kernel $\phi(x) = \sin(x)/ \pi$ for all $x \in (0, 2 \pi)$. As a consequence, we obtain the conclusion of the theorem.
\end{proof}

\subsection{Optimal function with Cauchy density }
In the previous section we saw that the sin function is optimal for minimizing $\int_\mathbb{R} (\phi(x)/x)^2\, dx$ subject to the constraints~\eqref{eq:normalized_constant}, namely, $\int_\mathbb{R}(\phi(x)/x)\,d x=1$ and $\int_0^{2\pi}\phi(x)\,d x=0$ and $\phi$ being a cyclic function on $(0,2\pi)$. That objective function stems from the upper bound~\eqref{eq:variance_upper_bound} on the variance of density estimator $\widehat{m}_{R}$.

In this section we demonstrate the reason in finding other integral theorems to Fourier's by finding optimal function $\phi$, also satisfying the constraints~\eqref{eq:normalized_constant}, which minimizes the leading term from the variance of $\widehat{m}_{R}$, which is given by: 
$$\int_\mathbb{R} \left(\frac{\phi(x)}{x}\right)^2\,m(x)\,d x$$
for some density function $m$ on $\mathbb{R}$. Since the above integral captures the leading term of the variance of $\widehat{m}_{R}$, it gives more precise asymptotic behavior of the variance than that of the term~\eqref{opt_optimistic}. Furthermore, as that integral involves the density function $m$, it indicates that the optimal kernel $\phi$ also depends on $m$. To illustrate our findings, we specifically consider the setting when $m$ is the Cauchy density; i.e., $m(x)=\pi^{-1}/(1+x^2)$ for all $x$,
which would be useful when modeling heavy tailed distributions and moreover we are able to solve the relevant equations.  
\begin{figure}[!t]
\begin{center}
\includegraphics[width=14cm,height=6.5cm]{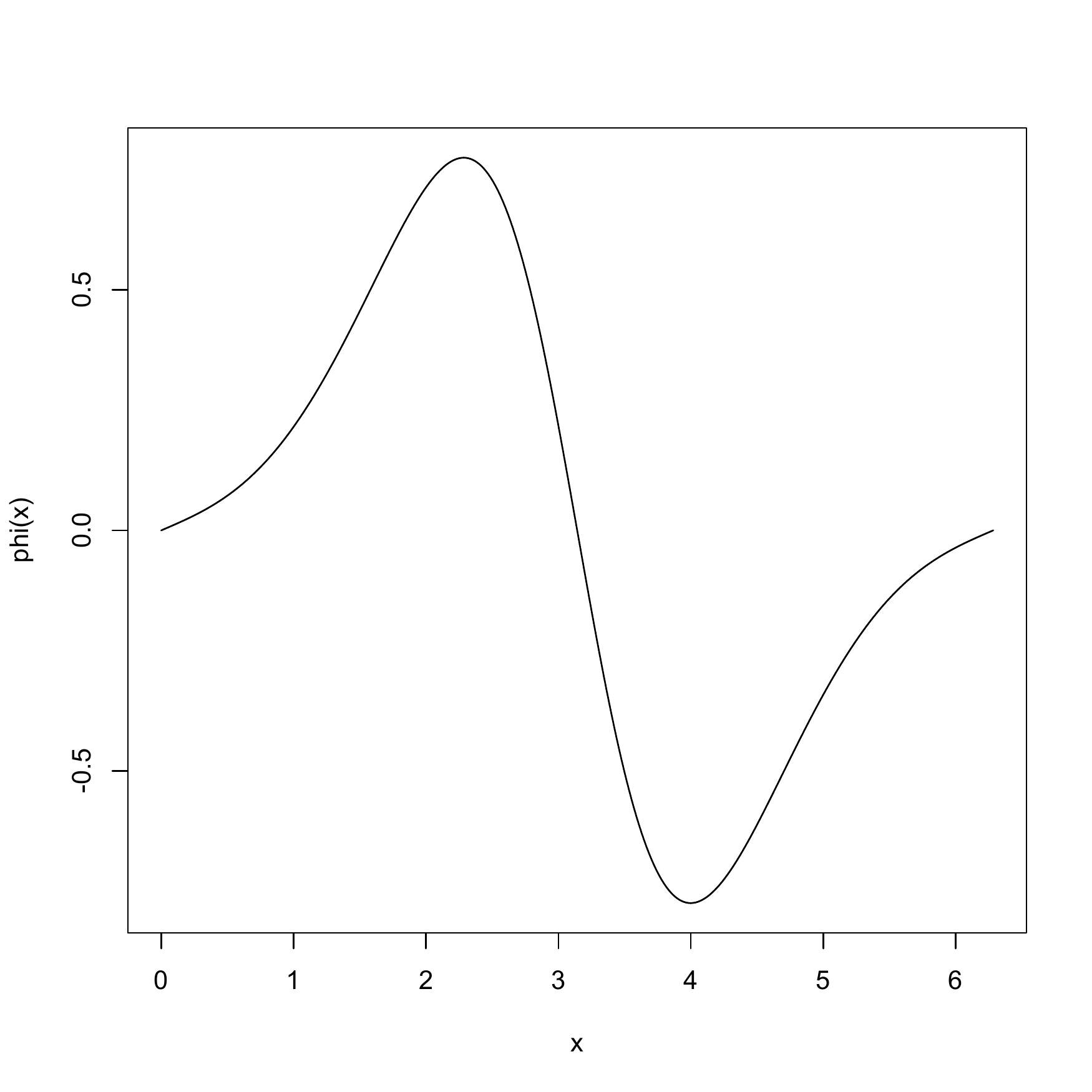}
\caption{Optimal function $\phi_{\text{Cauchy}}(x)$ for solving objective function~\eqref{opt_constrained} with respect to the Cauchy density function. As we can observe, it is different from the sin function from the Fourier integral theorem for solving objective function~\eqref{opt_optimistic}.}
\label{fig1}
\end{center}
\end{figure}

The proof idea for obtaining the optimal kernel $\phi$ is the same as in that of Theorem~\ref{theorem:optimal_function}. The only thing that is different is that we need to find a new function $\beta$, which we refer to as $\beta_{\text{Cauchy}}$; given now by:
$$\beta_{\text{Cauchy}}(x)=\sum_{m=-\infty}^\infty\frac{1}{(x+2\pi m)^2}\,\frac{1}{1+(x+2\pi m)^2}.$$
To find the closed-form expression of $\beta_{\text{Cauchy}}(x)$, we utilize contour integration from complex analysis (see for example \cite{Priest1985}); so consider
$$I_R=\int_{\gamma_R}\frac{\cot(z/2)\,dz}{(x+z)^2(1+(x+z)^2)},$$
where $\gamma_R$ is a circle in the complex plane of radius $R$ around the origin. The simple poles occur at $z=2\pi m$ for all integers $m$, giving a total residue of $2\beta(x)$, since the relevant coefficient in the Laurent expansion of $\cot(z/2)$ is 2; also at $z=x\pm i$ for which the residues are 
$-\cot\left(\half(x\pm i)\right)$. There is a double pole at $z=-x$ for which the residue is the first derivative of
$$f(z)=\frac{\cot(z/2)}{1+(x+z)^2}$$
evaluated at $z=-x$. From direct calculation, this term is $\half/\sin^2(x/2)$.

Now using the Cauchy residue theorem and noting that $I_R\to 0$ as $R\to\infty$, and expanding $\cot((x\pm i)/2)$, we obtain
$$\beta_{\text{Cauchy}}(x)=\qart\sin^{-2}(x/2)-\half\coth(1/2)\frac{1+\cot^2(x/2)}{\coth^2(1/2)+\cot^2(x/2)}.$$
As shown in the proof of Theorem~\ref{theorem:optimal_function}, the optimal function $\phi$ is of the form
\begin{align}
\phi_{\text{Cauchy}}(x)=\frac{\lambda_1+\lambda_2\alpha(x)}{\beta_{\text{Cauchy}}(x)}, \label{eq:new_optimal}
\end{align}
where recall that $\alpha(x)=\cot(x/2)$, with the $\lambda_{1}$, $\lambda_{2}$ values being now able to capture the coefficient of $\half$. Different from Theorem~\ref{theorem:optimal_function}, we do not have closed-form expressions for $\lambda_{1}$ and $\lambda_{2}$. However, numerical integration can be used to determine the $\lambda_{1}, \lambda_{2}$ values to meet the constraints~\eqref{eq:normalized_constant}, and this yields $\lambda_1=0$ and $\lambda_2=0.118$.
A picture of the optimal function $\phi_{\text{Cauchy}}$ is given in Figure~\ref{fig1}.


We note that such a result is not restricted to only the setting when the samples are generated from the Cauchy distribution; actually, when we have samples from a Gaussian distribution, using the kernel $\phi_{\text{Cauchy}}$ also yields slightly better variance than the sin kernel from the Fourier integral theorem. We leave a detailed investigation of the benefit of $\phi_{\text{Cauchy}}$ over the sin kernel for general settings of density estimation problem in the future work.


\begin{figure}[!t]
\begin{center}
\includegraphics[width=14cm,height=7cm]{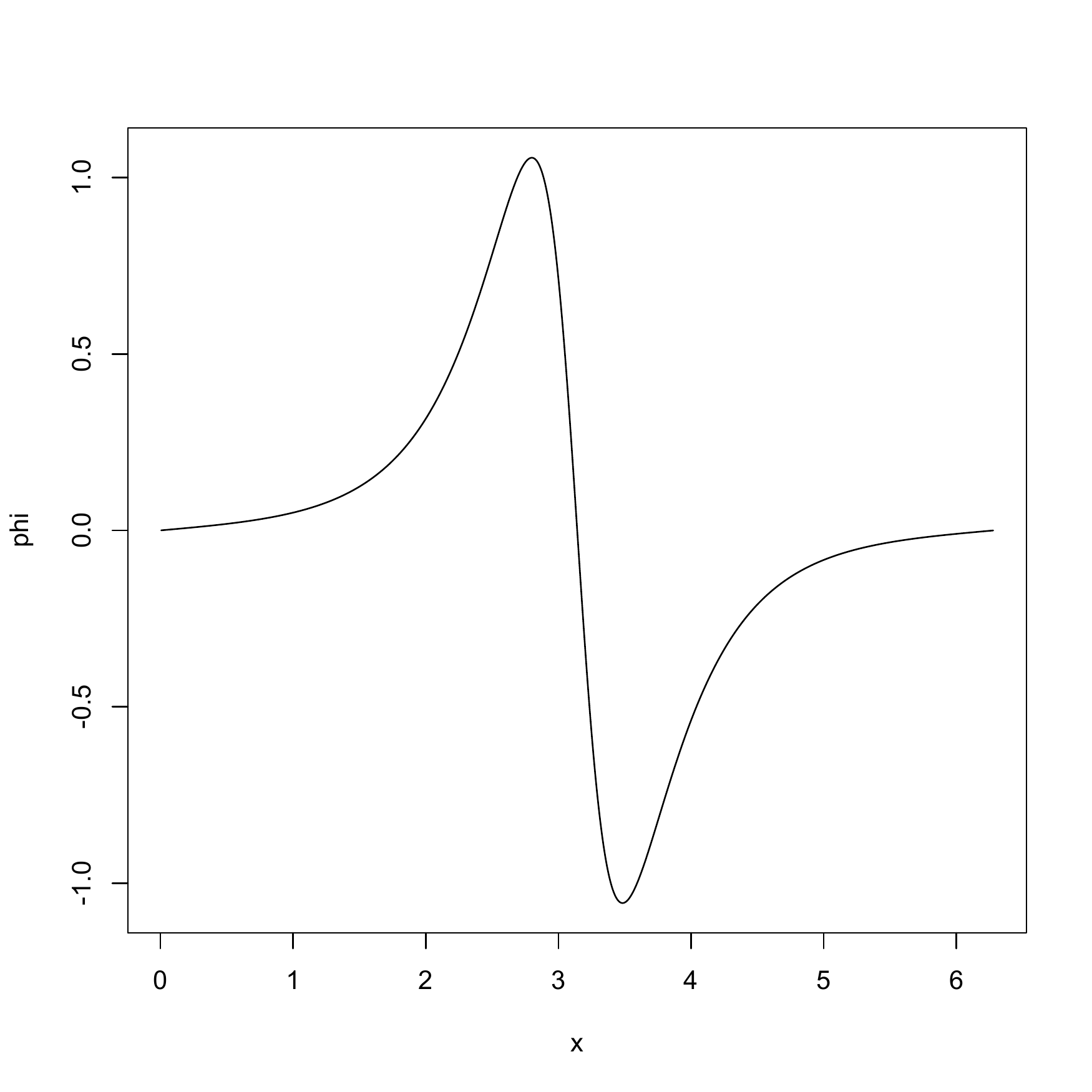}
\caption{The $\phi(x)$ for the standard Gaussian density.}
\label{fig3}
\end{center}
\end{figure}

For finding the optimal cyclic function for the general density function $m$ we would need to be able to find the function 
$$\sum_{k=-\infty}^{+\infty} \frac{m(x+2\pi k)}{(x+2\pi k)^2}.$$
This can always be tackled using contour integration though is not straightforward in many cases. Numerical solutions are obviously easy to get. For example, for the Gaussian density, the $\phi$ function (up to proportionality) appears in Fig.~\ref{fig3}, but as yet we have been unable to find an explicit solution.

\section{Discussion}
\label{sec:discussion}
In this paper we have introduced a general class of integral theorems. These provide natural Monte Carlo density estimators which can automatically preserve the dependence structure of a dataset. In the univariate density estimation setting, we demonstrate that the Fourier integral theorem is optimal when we minimize a square integral in equation~\eqref{opt_optimistic}; a term that indicates a good variance of the density estimator. 
To show the benefit of a general class of integral theorems, we also consider optimal kernels that minimize the term~\eqref{opt_constrained}, which provides a more precise nature of the variance of the kernel density estimator. Our study shows that the optimal kernel for alternative objective function are generally not the sin kernel.

Here we discuss a few future directions of research. First,  we have only obtained the optimal kernels in our general class of integral theorems for density estimation. It is also important to study optimal kernels for other statistical estimation tasks, such as nonparametric (modal) regression~\cite{Nadaraya_1964, Wass06, Tsy09} and mode clustering~\citep{Azzalini_2007, Chacon_2013, Chacon_2015}. Second, the work of~\cite{Thorp_2021} proposes using double Fourier transforms to approximate a nonparametric function that can capture both the correlation of words in each sequence and the correlation of sequences in natural language processing tasks. Given our study with the general integral theorems, it is of interest to investigate whether we can develop the general notion of double Fourier transforms in the similar way as we do for the integral theorems and whether the choice of double Fourier transforms is optimal for estimating the nonparametric function arising in natural language processing tasks. 
\section{Appendix}
\label{sec:Appendix}
In this Appendix, we give the proof of Theorem~\ref{theorem:approximation_Fourier}. To ease the presentation, the values of universal constants (e.g., $C$, $C_{1}$, $C_{2}$, $\bar{C}$ etc.) can change from line-to-line. For any $x \in \mathbb{R}^{d}$, we denote $x = (x_{1}, \ldots, x_{d})$. 
\subsection{Proof of Theorem~\ref{theorem:approximation_Fourier}}
We first prove the result of Theorem~\ref{theorem:approximation_Fourier} when $d = 1$. In particular, we would like to show that when the function $m \in \mathcal{T}^{K}(\mathbb{R})$, there exists a universal constant $C > 0$ such that we have
\begin{align*}
	\abss{m_{R}(y) - m(y)} \leq \frac{C}{R^{K}}.
\end{align*}
In fact, from the definition of $m_{R}(y)$ in equation~\eqref{eq:Fourier_approx}, we have
\begin{align*}
    \abss{m_{R}(y) - m(y)} = \abss{\frac{1}{\pi} \int_{\mathbb{R}} \frac{\sin(R(y-x))}{(y - x)} \parenth{m(x) - m(y)} dx}.
\end{align*}
For simplicity of the presentation, for any $y \in \mathbb{R}$ we write $g(x) = (m(x) - m(y))/(x - y)$ for all $x \in \mathbb{R}$. Then, we can rewrite the above equality as 
\begin{align*}
    \abss{m_{R}(y) - m(y)} & = \abss{\frac{1}{\pi} \int_{\mathbb{R}} \sin(R(y-x)) g(x) dx} \\
    & = \abss{\frac{1}{\pi} \sum_{k = -\infty}^{\infty} \int_{y + \frac{2\pi k}{R}}^{y + \frac{2\pi(k + 1)}{R}} \sin(R(y-x)) g(x) dx}.
\end{align*}
Invoking the change of variables $x = y + \frac{t + 2 \pi k}{R}$, the above equation becomes
\begin{align}
    \abss{m_{R}(y) - m(y)} & = \abss{\frac{1}{\pi R} \sum_{k = -\infty}^{\infty} \int_{[0, 2\pi)} \sin(t) \cdot g \parenth{y + \frac{t + 2 \pi k}{R}}dt}. \label{eq:key_equation}
\end{align}
Since $m \in \mathcal{T}^{K}(\mathbb{R})$, the function $g$ is differentiable up to the $K$-th order. Therefore, using a Taylor expansion up to the $K$-th order, leads to 
\begin{align*}
    g \parenth{y + \frac{t + 2 \pi k}{R}} & = \sum_{\alpha \leq K - 1} \frac{1}{R^{\alpha}} \frac{t^{\alpha}}{\alpha!} g^{(\alpha)} \parenth{y + \frac{2 \pi k}{R}} \\
    & \hspace{4 em} + \frac{t^{K}}{R^{K} (K - 1)!} \int_{0}^{1} (1 - \xi)^{K - 1} g^{(K)} \parenth{y + \frac{2 \pi k}{R} + \frac{\xi t}{R}} d \xi.
\end{align*} 
Plugging the above Taylor expansion into equation~\eqref{eq:key_equation}, we have
\begin{align}
    \abss{m_{R}(y) - m(y)} = \abss{\frac{1}{\pi R} \sum_{\ell = 0}^{K} A_{\ell}} \leq \frac{1}{\pi R} \sum_{\ell = 0}^{K} \abss{A_{\ell}}, \label{eq:key_inequality_Fourier}
\end{align}
where, for $\ell \in \{0,1, \ldots, K - 1\}$, we define
\begin{align*}
    A_{\ell} & =  \frac{1}{R^{\ell}} \int_{[0, 2\pi)} \parenth{\frac{t^{\ell} \sin(t)}{\ell !}} dt \parenth{ \sum_{k = -\infty}^{\infty} g^{(\ell)} \parenth{y + \frac{2 \pi k}{R}}}, \ \text{and} \\
    A_{K} & = \sum_{k = -\infty}^{\infty} \int_{[0, 2\pi)} \sin(t) \parenth{ \frac{t^{K}}{R^{K}(K - 1)!} \int_{0}^{1} (1 - \xi)^{K - 1} g^{(K)} \parenth{y + \frac{2 \pi k}{R} + \frac{\xi t}{R}} d \xi} dt.
\end{align*}
We now find a bound for $\abss{A_{\ell}}$ for $\ell \in \{0,1,\ldots,K - 1\}$; we will demonstrate that
\begin{align}
    \abss{\sum_{k = -\infty}^{\infty} g^{(\ell)} \parenth{y + \frac{2 \pi k}{R}}} \leq \frac{C}{R^{K - (\ell + 1)}}, \quad \text{for all} \ \ell \in \{0,1,\ldots,K - 1\} \label{eq:inequality_first}
\end{align}
where $C$ is some universal constant. To obtain these bounds, we will use an inductive argument on $\ell$. We first start with $\ell = K - 1$. In fact, we have
\begin{align*}
    & \hspace{-5 em} \biggr|\sum_{k = -\infty}^{\infty} g^{(K - 1)} \parenth{y + \frac{2 \pi k}{R}} \frac{(2\pi)}{R} - \int_{\mathbb{R}} g^{(K - 1)}(x) dx \biggr| \\
    & \hspace{3 em} = \biggr|\sum_{k = -\infty}^{\infty} \int_{y + \frac{2\pi k}{R}}^{y + \frac{2\pi(k + 1)}{R}}  \biggr(g^{(K - 1)}(x) - g^{(K - 1)} \parenth{y + \frac{2 \pi k}{R}}\biggr)  dx \biggr| \\
    & \hspace{3 em} \leq \sum_{k = -\infty}^{\infty} \int_{y + \frac{2\pi k}{R}}^{y + \frac{2\pi(k + 1)}{R}} \abss{g^{(K - 1)}(x) - g^{(K - 1)} \parenth{y + \frac{2 \pi k}{R}}} dx.
\end{align*}
An application of Taylor expansion leads to
\begin{align*}
    g^{(K - 1)}(x) & = g^{(K - 1)} \parenth{y + \frac{2 \pi k}{R}} +  \parenth{x - y - \frac{2\pi k}{R}} \int_{0}^{1} g^{(K)} \parenth{(1 - \xi) \parenth{y + \frac{2 \pi k}{R}} + \xi x } d \xi.
\end{align*}
Now for any $\xi \in [0, 1]$ and $x \in \brackets{y + \frac{2\pi k}{R}, y + \frac{2 \pi (k + 1)}{R}}$, we have
\begin{align*}
    \abss{g^{(K)} \parenth{(1 - \xi) \parenth{y + \frac{2 \pi k}{R}} + \xi x }} \leq  \sup_{t \in \brackets{y + \frac{2\pi k}{R}, y + \frac{2 \pi (k + 1)}{R}}} \abss{g^{(K)}(t)}.
\end{align*}
Collecting the above results, we find that 
\begin{align*}
    & \hspace{- 3 em} \sum_{k = -\infty}^{\infty} \int_{y + \frac{2\pi k}{R}}^{y + \frac{2\pi(k + 1)}{R}} \abss{g^{(K - 1)}(x) - g^{(K - 1)} \parenth{y + \frac{2 \pi k}{R}}} dx \\
    & \leq \sum_{k = -\infty}^{\infty} \parenth{\int_{y + \frac{2\pi k}{R}}^{y + \frac{2\pi(k + 1)}{R}} \parenth{x - y - \frac{2\pi k}{R}} dx} \sup_{t \in \brackets{y + \frac{2\pi k}{R}, y + \frac{2 \pi (k + 1)}{R}}} \abss{g^{(K)}(t)} \\
    & = \frac{2 \pi^{2}}{R^{2}} \sum_{k = -\infty}^{\infty} \sup_{t \in \brackets{y + \frac{2\pi k}{R}, y + \frac{2 \pi (k + 1)}{R}}} \abss{g^{(K)}(t)}.
\end{align*}
Using a Riemann sums approximating integrals theorem, we have
\begin{align*}
	\lim_{R \to \infty} \sum_{k = -\infty}^{\infty} \sup_{t \in \brackets{y + \frac{2\pi k}{R}, y + \frac{2 \pi (k + 1)}{R}}} \abss{g^{(K)}(t)} \frac{2\pi}{R} = \int_{\mathbb{R}} \abss{g^{(K)}(x)}dx < \infty,
\end{align*}
where the finite value of the integral is due to the assumption that $m \in \mathcal{T}^{K}(\mathbb{R})$. Furthermore, the above limit is uniform in terms of $y$ as $g^{(K)}$ is uniformly continuous. Collecting the above results, there exists a universal constant $C$ such that as long as $R \geq C$, the following inequality holds:
\begin{align*}
    \frac{2 \pi^{2}}{R^{2}} \sum_{k = -\infty}^{\infty} \sup_{t \in \brackets{y + \frac{2\pi k}{R}, y + \frac{2 \pi (k + 1)}{R}}} \abss{g^{(K)}(t)} \leq \frac{C_{1}}{R}
\end{align*}
where $C_{1}$ is some universal constant. Combining all of the previous results, we obtain 
\begin{align*}
    \biggr|\sum_{k = -\infty}^{\infty} g^{(K - 1)} \parenth{y + \frac{2 \pi k}{R}} \frac{(2\pi)}{R} - \int_{\mathbb{R}} g^{(K - 1)}(x) dx \biggr|  \leq \frac{C_{1}}{R}.
\end{align*}
Since $m \in \mathcal{T}^{K}(\mathbb{R})$, using integration by parts, we get
\begin{align*}
    \int_{\mathbb{R}} g^{(K - 1)}(x) dx= 0.
\end{align*}
Therefore, we obtain the conclusion of equation~\eqref{eq:inequality_first} when $\ell = K - 1$. 

Now assume that the conclusion of equation~\eqref{eq:inequality_first} holds for $1 \leq \ell \leq K - 1$. We will prove that the conclusion also holds for $\ell - 1$. With a similar argument to the setting $\ell = K - 1$, we  obtain 
\begin{align}
	    & \hspace{-5 em} \biggr|\sum_{k = -\infty}^{\infty} g^{(\ell)} \parenth{y + \frac{2 \pi k}{R}} \frac{(2\pi)}{R} - \int_{\mathbb{R}} g^{(\ell)}(x) dx \biggr| \nonumber \\
	    & \hspace{3 em} = \abss{\sum_{k = -\infty}^{\infty} \int_{y + \frac{2\pi k}{R}}^{y + \frac{2\pi(k + 1)}{R}} \parenth{g^{(\ell)}(x) - g^{(\ell)} \parenth{y + \frac{2 \pi k}{R}}} dx}. \label{eq:inductive_argument}
\end{align}
Using a Taylor expansion, we have
\begin{align*}
    g^{(\ell)}(x) & =  g^{(\ell)} \parenth{y + \frac{2 \pi k}{R}} + \sum_{\alpha \leq K - 1 - \ell} \frac{\parenth{x - y - \frac{2\pi k_{j}}{R}}^{\alpha}}{\alpha!} g^{(\ell + \alpha)} \parenth{y + \frac{2 \pi k}{R}} \\
    & + \frac{\parenth{x - y - \frac{2\pi k}{R}}^{K - \ell}}{(K - \ell - 1)!} \int_{0}^{1} (1 - \xi)^{K - \ell - 1} g^{(K)} \parenth{(1 - \xi) \parenth{y + \frac{2 \pi k}{R}} + \xi x } d \xi.
\end{align*}
Plugging the above Taylor expansion into equation~\eqref{eq:inductive_argument}, we find that
\begin{align*}
   \sum_{k = -\infty}^{\infty} \int_{y + \frac{2\pi k}{R}}^{y + \frac{2\pi(k + 1)}{R}} \abss{g^{(\ell)}(x) - g^{(\ell)} \parenth{y + \frac{2 \pi k}{R}}} dx \leq S_{1} + S_{2},
\end{align*}
where $S_{1}$ and $S_{2}$ are defined as follows:
\begin{align*}
    S_{1} & = \sum_{\alpha \leq K - 1 - \ell} \biggr|\sum_{k = -\infty}^{\infty} \parenth{\int_{y + \frac{2\pi k}{R}}^{y + \frac{2\pi(k + 1)}{R}} \frac{\parenth{x - y - \frac{2\pi k_{j}}{R}}^{\alpha}}{\alpha!} dx} g^{(\ell + \alpha)} \parenth{y + \frac{2 \pi k}{R}} \biggr| \\
    & = \sum_{\alpha \leq K - 1 - \ell}  \biggr|\sum_{k = -\infty}^{\infty} \frac{(2\pi)^{\alpha + 1}}{R^{\alpha + 1} (\alpha+1)!} g^{(\ell + \alpha)} \parenth{y + \frac{2 \pi k}{R}} \biggr|;
\end{align*}
\begin{align*}
    S_{2} & = \biggr|\sum_{k = -\infty}^{\infty} \biggr(\int_{y + \frac{2\pi k}{R}}^{y + \frac{2\pi(k + 1)}{R}}\frac{\parenth{x - y - \frac{2\pi k}{R}}^{K - \ell}}{(K - \ell - 1)!} \\
    &  \hspace{6 em} \times \int_{0}^{1} (1 - \xi)^{K - \ell - 1} g^{(K)} \parenth{(1 - \xi) \parenth{y + \frac{2 \pi k}{R}} + \xi x } d \xi \biggr) dx \biggr| \\
    & \leq \sum_{k = -\infty}^{\infty} \int_{y + \frac{2\pi k}{R}}^{y + \frac{2\pi(k + 1)}{R}} \int_{0}^{1} \frac{\parenth{x - y - \frac{2\pi k}{R}}^{K - \ell}}{(K - \ell - 1)!} (1 - \xi)^{K - \ell - 1} \sup_{t \in \brackets{y + \frac{2\pi k}{R}, y + \frac{2 \pi (k + 1)}{R}}} \abss{g^{(K)}(t)} d\xi dx \\
    & = \sum_{k = -\infty}^{\infty}(K - \ell) \frac{(2\pi)^{K - \ell + 1}}{R^{K - \ell + 1} (K - \ell+1)!} \sup_{t \in \brackets{y + \frac{2\pi k}{R}, y + \frac{2 \pi (k + 1)}{R}}} \abss{g^{(K)}(t)}.
\end{align*}
An application of the triangle inequality and the hypothesis of the induction argument shows that
\begin{align*}
    S_{1} & \leq \sum_{\alpha \leq K - 1 - \ell} \frac{(2\pi)^{\alpha + 1}}{R^{\alpha + 1} (\alpha+1)!} \abss{\sum_{k = -\infty}^{\infty} g^{(\ell + \alpha)} \parenth{y + \frac{2 \pi k}{R}}} \leq \frac{C_{2}}{R^{K - \ell}}
\end{align*}
where $C_{2}$ is some universal constant.

Similar to the setting $\ell = K - 1$, an application of the Riemann sums approximating integrals theorem leads to
\begin{align*}
	S_{2} & \leq (K - \ell) \frac{(2\pi)^{K - \ell + 1}}{R^{K - \ell + 1} (K - \ell+1)!} \sum_{k = -\infty}^{\infty}  \sup_{t \in \brackets{y + \frac{2\pi k}{R}, y + \frac{2 \pi (k + 1)}{R}}} \abss{g^{(K)}(t)} \leq \frac{C_{3}}{R^{K - \ell}},
\end{align*}
when $R$ is sufficiently large where $C_{3}$ is some universal constant. Putting all the above results together, as long as $R \geq C$ we have
\begin{align*}
    \biggr|\sum_{k = -\infty}^{\infty} g^{(\ell)} \parenth{y + \frac{2 \pi k}{R}} \frac{(2\pi)}{R} - \int_{\mathbb{R}} g^{(\ell)}(x) dx \biggr| \leq \frac{\bar{C}}{R^{K - \ell}}
\end{align*}
for some universal constant $\bar{C}$. As $\int_{\mathbb{R}} g^{(\ell)}(x) dx = 0$,
the above inequality leads to the conclusion of equation~\eqref{eq:inequality_first} for $1 \leq \ell \leq K - 1$. As a consequence, we obtain the conclusion of equation~\eqref{eq:inequality_first} for all $\ell \in \{0, 1, \ldots, K - 1\}$. 

Given equation~\eqref{eq:inequality_first}, an application of the triangle inequality leads to
\begin{align}
    \abss{A_{\ell}} \leq \frac{1}{R^{\ell}} \abss{\int_{[0, 2\pi)} \frac{t^{\ell} \sin(t)}{\ell!} dt} \abss{ \sum_{k = -\infty}^{\infty} g^{(\ell)} \parenth{y + \frac{2 \pi k}{R}}} \leq \frac{\bar{C}}{R^{K - 1}} \label{eq:bound_Al}
\end{align} 
for all $0 \leq \ell \leq K - 1$. 

We now find a bound for $\abss{A_{K}}$. A direct application of the triangle inequality leads to the following bound of $\abss{A_{K}}$:
\begin{align*}
	\abss{A_{K}} \leq \int_{[0, 2\pi)} \frac{\abss{\sin(t) t^{K}}}{R^{K}(K - 1)!} dt \parenth{\sum_{k = -\infty}^{\infty} \int_{0}^{1} (1 - \xi)^{K - 1} \abss{g^{(K)} \parenth{y + \frac{2 \pi k}{R} + \frac{\xi t}{R}}} d \xi}.
\end{align*}
For any $\xi \in [0,1]$ and $t \in [0, 2\pi)$, we have
\begin{align*}
    \abss{g^{(K)} \parenth{y + \frac{2 \pi k}{R} + \frac{\xi t}{R}}} \leq \sup_{x \in \brackets{y + \frac{2\pi k}{R}, y + \frac{2\pi (k + 1)}{R}}} \abss{g^{(K)}(x)}.
\end{align*}
Putting the above inequalities together, we find that
\begin{align*}
    \abss{A_{K}} & \leq \int_{[0, 2\pi)} \frac{\abss{\sin(t) t^{K}}}{R^{K} K!} dt \parenth{\sum_{k = -\infty}^{\infty} \sup_{x \in \brackets{y + \frac{2\pi k}{R}, y + \frac{2\pi (k + 1)}{R}}} \abss{g^{(K)}(x)}}.
\end{align*}
From the Riemann sums approximating integrals theorem, we obtain
\begin{align*}
    \frac{(2\pi)}{R^{1 - K}} \abss{A_{K}} \leq C_{1} \int_{[0, 2\pi)} \frac{\abss{\sin(t) t^{K}}}{R^{K} K!} dt,
\end{align*}
where $C_{1}$ is some universal constant. Collecting the above results, we conclude that 
\begin{align}
    |A_{K}| \leq \frac{\bar{C}}{R^{K - 1}} \label{eq:bound_AK}
\end{align}
for some constant $\bar{C}$. 
Putting the bounds~\eqref{eq:bound_Al} and~\eqref{eq:bound_AK} into equation~\eqref{eq:key_inequality_Fourier}, we obtain the conclusion of the theorem when $d = 1$.

\vspace{0.5 em}
We now provide the proof of Theorem~\ref{theorem:approximation_Fourier} for general dimension $d$.

\noindent
When $m(x) = \sum_{j = 1}^{d} m_{j}(x_{j})$ for any $x = (x_{1}, \ldots, x_{d})$: From the definition of $m_{R}(y)$ in equation~\eqref{eq:Fourier_approx}, we have
\begin{align*}
    \abss{m_{R}(y) - m(y)} & = \abss{\frac{1}{\pi^{d}} \int_{\mathbb{R}^{d}} \prod_{j = 1}^{d} \frac{\sin(R(y_j-x_j))}{(y_{j} - x_{j})} \parenth{m(x) - m(y)} dx} \\
    & = \abss{\frac{1}{\pi^{d}} \int_{\mathbb{R}^{d}} \prod_{j = 1}^{d} \frac{\sin(R(y_j-x_j))}{(y_{j} - x_{j})} \parenth{\sum_{j = 1}^{d} m_{j}(x_{j}) - \sum_{j = 1}^{d} m_{j}(y_{j})} dx} \\
    & \leq \sum_{j = 1}^{d} \abss{\frac{1}{\pi} \int_{\mathbb{R}} \frac{\sin(R(y_j-x_j))}{(y_{j} - x_{j})} \parenth{m_{j}(x_{j}) - m_{j}(y_{j})} dx_{j}}.
\end{align*}
Since $m_{j} \in \mathcal{T}^{K_{j}}(\mathbb{R})$ for $1 \leq j \leq d$, an application of the result of Theorem~\ref{theorem:approximation_Fourier} when $d = 1$ leads to 
\begin{align*}
	\abss{\frac{1}{\pi} \int_{\mathbb{R}} \frac{\sin(R(y_j-x_j))}{(y_{j} - x_{j})} \parenth{m_{j}(x_{j}) - m_{j}(y_{j})} dx_{j}} \leq \frac{C_{j}}{R^{K_{j}}}
\end{align*}
where $C_{j}$ are universal constants. Putting the above results together, we obtain the conclusion of the theorem when $m$ is the summation of the functions $m_{1}, m_{2}, \ldots, m_{d}$.

\vspace{0.5 em}
\noindent
When $m(x) = \prod_{j = 1}^{d} m_{j}(x_{j})$ for any $x = (x_{1}, \ldots, x_{d})$: Similar to the argument when $m$ is the summation of $m_{1}, m_{2}, \ldots, m_{d}$, we have
\begin{align*}
    \abss{m_{R}(y) - m(y)} & = \abss{\frac{1}{\pi^{d}} \int_{\mathbb{R}^{d}} \prod_{j = 1}^{d} \frac{\sin(R(y_j-x_j))}{(y_{j} - x_{j})} \parenth{\prod_{j = 1}^{d} m_{j}(x_{j}) - \prod_{j = 1}^{d} m_{j}(y_{j})} dx} \\
    & = \biggr|\frac{1}{\pi^{d}} \int_{\mathbb{R}^{d}} \prod_{j = 1}^{d} \frac{\sin(R(y_j-x_j))}{(y_{j} - x_{j})} \biggr(\sum_{\ell = 0}^{d - 1} \prod_{j = 1}^{\ell} m_{j}(y_{j}) \prod_{j = \ell + 1}^{d} m_{j}(x_{j}) \\
    & \hspace{14 em} - \prod_{j = 1}^{\ell + 1} m_{j}(y_{j}) \prod_{j = \ell + 2}^{d} m_{j}(x_{j})\biggr) dx\biggr| \\
    & \leq \sum_{\ell = 0}^{d - 1} \biggr|\frac{1}{\pi^{d}} \int_{\mathbb{R}^{d}} \prod_{j = 1}^{d} \frac{\sin(R(y_j-x_j))}{(y_{j} - x_{j})} \biggr(\prod_{j = 1}^{\ell} m_{j}(y_{j}) \prod_{j = \ell + 1}^{d} m_{j}(x_{j}) \\
    & \hspace{14 em} - \prod_{j = 1}^{\ell + 1} m_{j}(y_{j}) \prod_{j = \ell + 2}^{d} m_{j}(x_{j})\biggr) dx \biggr| \\
    & \leq C \sum_{\ell = 0}^{d - 1} \abss{\frac{1}{\pi} \int_{\mathbb{R}} \frac{\sin(R(y_{\ell + 1}-x_{\ell + 1}))}{(y_{\ell + 1} - x_{\ell + 1})} \parenth{m_{\ell + 1}(x_{\ell + 1}) - m_{\ell + 1}(y_{\ell + 1})} dx_{\ell + 1}}
\end{align*}
where $C$ is some universal constant. Using the above bound and the result in one dimension of Theorem~\ref{theorem:approximation_Fourier} for $m_{1}, m_{2}, \ldots, m_{d}$, we obtain the conclusion of the theorem when $m$ is the product of these functions.
\bibliographystyle{plainnat}
\bibliography{Nhat}

\begin{thebibliography}{31}
\providecommand{\natexlab}[1]{#1}
\providecommand{\url}[1]{\texttt{#1}}
\expandafter\ifx\csname urlstyle\endcsname\relax
  \providecommand{\doi}[1]{doi: #1}\else
  \providecommand{\doi}{doi: \begingroup \urlstyle{rm}\Url}\fi

\bibitem[Azzalini and Torelli(2007)]{Azzalini_2007}
A.~Azzalini and N.~Torelli.
\newblock Clustering via nonparametric density estimation.
\newblock \emph{Statistics and Computing}, 17:\penalty0 71--80, 2007.

\bibitem[Bochner(1959)]{Bochner_1959}
S.~Bochner.
\newblock \emph{Lectures on Fourier Integrals}.
\newblock Princeton University Press, 1959.

\bibitem[Botev et~al.(2010)Botev, Grotowski, and Kroese]{Botev_Diffusion}
Z.~I. Botev, J.~F. Grotowski, and D.~P. Kroese.
\newblock Kernel density estimation via diffusion.
\newblock \emph{Annals of Statistics}, 38:\penalty0 2916--2957, 2010.

\bibitem[Chac\'{o}n(2015)]{Chacon_2015}
J.~E. Chac\'{o}n.
\newblock A population background for nonparametric density-based clustering.
\newblock \emph{Statistical Science}, 30:\penalty0 518--532, 2015.

\bibitem[Chac\'{o}n and Duong(2013)]{Chacon_2013}
J.~E. Chac\'{o}n and T.~Duong.
\newblock Data-driven density derivative estimation, with applications to
  nonparametric clustering and bump hunting.
\newblock \emph{Electronic Journal of Statistics}, 7:\penalty0 499--532, 2013.

\bibitem[Chacon and Duong(2018)]{Chacon18}
J.E. Chacon and T.~Duong.
\newblock \emph{Multivariate Kernel Smoothing and its Applications}.
\newblock CRC Press, 2018.

\bibitem[Davis(1975)]{Davis75}
K.B. Davis.
\newblock Mean square error properties of density estimates.
\newblock \emph{Annals of Statistics}, 3:\penalty0 1025--1030, 1975.

\bibitem[Dempster et~al.(1997)Dempster, Laird, and Rubin]{Rubin-1977}
A.~P. Dempster, N.~M. Laird, and D.~B. Rubin.
\newblock Maximum likelihood from incomplete data via the {EM} algorithm.
\newblock \emph{Journal of the Royal Statistical Society: Series B (Statistical
  Methodology)}, 39:\penalty0 1--38, 1997.

\bibitem[Epanechnikov(1969)]{Epanechnikov_1969}
V.~A. Epanechnikov.
\newblock Non-parametric estimation of a multivariate probability density.
\newblock \emph{Theory of Probability \& Its Applications}, 14:\penalty0
  153--158, 1969.

\bibitem[Fowler(1921)]{Fowler1921}
R.~H. Fowler.
\newblock A simple extension of {F}ourier's integral theorem and some physical
  applications, in particular to the theory of quanta.
\newblock \emph{Proceedings of the Royal Society of London, Series A},
  99:\penalty0 462--471, 1921.

\bibitem[Gin\'{e} and Nickl(2010)]{Nickl_Confidence}
E.~Gin\'{e} and R.~Nickl.
\newblock Confidence bands in density estimation.
\newblock \emph{Annals of Statistics}, 38:\penalty0 1122--1170, 2010.

\bibitem[Gy\"{o}rfi et~al.(1985)Gy\"{o}rfi, Devroye, , and
  Gyorf]{Devroye_Nonparametric}
L.~Gy\"{o}rfi, L.~Devroye, , and L.~Gyorf.
\newblock \emph{Nonparametric Density Estimation: the L1 View}.
\newblock John Wiley \& Sons, New York, 1985.

\bibitem[Ho and Walker(2021)]{Ho21}
N.~Ho and S.G. Walker.
\newblock Multivariate smoothing via the {F}ourier integral theorem and
  {F}ourier kernel.
\newblock \emph{Under review, Journal of Machine Learning Research}, 2021.

\bibitem[Jiang(2017)]{Jiang_Uniform}
H.~Jiang.
\newblock Uniform convergence rates for kernel density estimation.
\newblock In \emph{ICML}, 2017.

\bibitem[Lee-Thorp et~al.(2021)Lee-Thorp, Ainslie, Eckstein, and
  Onta\"{n}\'{o}n]{Thorp_2021}
J.~Lee-Thorp, J.~Ainslie, I.~Eckstein, and S.~Onta\"{n}\'{o}n.
\newblock Fnet: Mixing tokens with {F}ourier transforms.
\newblock \emph{arXiv preprint}, 2021.

\bibitem[Mueller(1984)]{Mueller1984}
H-G. Mueller.
\newblock Smooth optimum kernel estimators of densities, regression curve and
  modes.
\newblock \emph{Annals of Statistics}, 12:\penalty0 766--774, 1984.

\bibitem[Nadaraya(1964)]{Nadaraya_1964}
E.~A. Nadaraya.
\newblock On estimating regression.
\newblock \emph{Theory of Probability \& Its Applications}, 9:\penalty0
  141--142, 1964.

\bibitem[Parzen(1962)]{Parzen62}
E.~Parzen.
\newblock On estimation of a probability density function and mode.
\newblock \emph{Annals of Mathematical Statistics}, 33:\penalty0 1065--1076,
  1962.

\bibitem[Priestley(1985)]{Priest1985}
H.~A. Priestley.
\newblock \emph{Introduction to Complex Analysis}.
\newblock Clarendon Press, Oxford, 1985.

\bibitem[Rosenblatt(1956)]{Rosen1956}
M.~Rosenblatt.
\newblock Remarks on some nonparametric estimates of a density function.
\newblock \emph{Annals of Mathematical Statistics}, 27:\penalty0 832--837,
  1956.

\bibitem[Rotiroti and Walker(2022)]{roti22}
F.~Rotiroti and S.G. Walker.
\newblock Computing marginal likelihoods via the fourier integral theorem and
  pointwise estimation of posterior densities.
\newblock \emph{Under revision for Statistics and Computing}, 2022.

\bibitem[Staniswalis et~al.(1993)Staniswalis, Messer, and Finston]{Stanis93}
J.G. Staniswalis, K.~Messer, and D.R. Finston.
\newblock Kernel estimators for multivariate regression.
\newblock \emph{Journal of Nonparametric Statistics}, 3:\penalty0 103--121,
  1993.

\bibitem[Terrell and Scott(1992)]{Scott_Variable}
G.~R. Terrell and D.~W. Scott.
\newblock Variable kernel density estimation.
\newblock \emph{Annals of Statistics}, 20:\penalty0 1236--1265, 1992.

\bibitem[Tsybakov(2009)]{Tsy09}
A.~Tsybakov.
\newblock \emph{Introduction to Nonparametric Estimation}.
\newblock Springer, 2009.

\bibitem[Wand(1992)]{Wand92}
M.P. Wand.
\newblock Error analysis for general multivariate kernel estimators.
\newblock \emph{Journal of Nonparametric Statistics}, 2:\penalty0 1--15, 1992.

\bibitem[Wand(1994)]{Wand94}
M.P. Wand.
\newblock Fast computation of multivariate kernel estimators.
\newblock \emph{Journal of Computational and Graphical Statistics}, 3:\penalty0
  433--445, 1994.

\bibitem[Wand and Jones(1993)]{WJones93}
M.P. Wand and M.C. Jones.
\newblock Comparison of smoothing parameterizations in bivariate kernel density
  estimation.
\newblock \emph{Journal of the American Statistical Association}, 88:\penalty0
  520--528, 1993.

\bibitem[Wasserman(2006)]{Wass06}
L.~Wasserman.
\newblock \emph{All of Nonparametric Statistics}.
\newblock Springer, 2006.

\bibitem[Wiener(1933)]{Wiener33}
N.~Wiener.
\newblock \emph{The Fourier Integral and Certain of its Applications}.
\newblock Cambridge University Press, 1933.

\bibitem[Yakowitz(1985)]{Yakowitz_1985}
S.~J. Yakowitz.
\newblock Nonparametric density estimation, prediction, and regression for
  {M}arkov sequences.
\newblock \emph{Journal of the American Statistical Association}, 80:\penalty0
  215--221, 1985.

\bibitem[Young(1969)]{Young1969}
L.~C. Young.
\newblock \emph{Lecture on the Calculus of Variations and Optimal Control
  Theory}.
\newblock AMS Chelsea Publishing, 1969.

\end{thebibliography}
\end{document}